\documentclass{PoS}

\newcommand{\tc}{T_c}
\newcommand{\beq}{\begin{equation}}
\newcommand{\eeq}{\end{equation}}
\newcommand{\stbo}{{\scriptscriptstyle SB}}

\title{Continuum Thermodynamics of the SU(N) Gauge Theory}

\ShortTitle{Continuum Thermodynamics of the SU(N) Gauge Theory}

\author{\speaker{Saumen Datta} and Sourendu Gupta \\
        Tata Institute of Fundamental Research\\
        E-mail: \email{saumen@theory.tifr.res.in,sgupta@tifr.res.in}}

\abstract{The thermodynamics of the deconfined phase of the SU(N) gauge theory 
is studied. Careful study is made of the approach to the continuum limit.
The latent heat of the deconfinement transition is studied, for the theories
with 3, 4 and 6 colors. Continuum estimates of various thermodynamic 
quantities are studied, and the approach to conformality investigated. 
The bulk thermodynamic quantities at different N are compared, to 
investigate the validity of 'tHooft scaling at these values of N.}

\FullConference{The XXVIII International Symposium on Lattice Filed Theory\\
		 June 14-19,2010\\
		 Villasimius, Sardinia Italy}

\begin{document}

\section{Introduction}
\label{sec.intro}
Interest in SU(N) gauge theories with large N began with the
pioneering studies of \cite{thooft}, who showed that
in the limit N $\to\infty$ and the gauge 
coupling, $g\to0$, with the 't Hooft coupling $\lambda=g^2$N fixed,
one gets a non-trivial but simplified theory. Many qualitative
features of hadron physics can be explained by appealing to this limit. 

The theory with an infinite number of colors has also been used to
understand various facets of the phase diagram of strongly interacting
matter. At very high temperatures, strongly interacting matter is known to
exist in a deconfined, chirally symmetric state. The nature of the
transition to this state is quite sensitive to the quark sector.
For infinitely massive quarks, one has a first order deconfinement 
transition for N $> 2$. For two massless quarks, on the other hand, 
one has a second order, chiral symmetry restoring transition. If a
small mass is given to the quarks, one expects a crossover, 
and a critical point at some nonzero baryon density. See
Ref. \cite{thermo} for a detailed discussion of the nature of the
transition. A rich phase
structure has also been predicted for the large baryon density
regime. In particular, a chirally symmetric, confined phase has been
predicted, using the 't Hooft limit \cite{quarkonic}. 
 
For phase diagrams based on arguments about the theory with an
infinite number of colors to be relevant for the theory with three
colors, it is important to explore the validity of the large N
arguments for N=3. Analogies have also been drawn between the high
temperature phase of QCD and the solvable, conformal ${\mathcal N}$ =4
supersymmetric SU(N $\to \infty$) theory \cite{gubser}. Lattice studies of the
theory with moderate values of N can provide one intermediate 
step in this connection, by giving an estimate of the size of the
$1/{\rm N}^2$ corrections.

In this report, we present results for a study of the thermodynamics 
of the SU(N) 
gauge theories at finite temperatures for N = 4,6, and combine them
with results for N=3 to get an estimate of the applicability of the 
large N arguments for the theory with three colors. For N=3, we
have used the simulation results of Ref. \cite{boyd}, supplementing
them with new simulations where necessary (in particular, for the
measurement of the latent heat), and analyzed them using the same
techniques as used for the N=4 and 6 theories.
For estimating the thermodynamic quantities, we use the standard
methods \cite{boyd}. The pressure, p, and $\Delta = \epsilon-3p$, where 
$\epsilon$ is the energy density, are calculated from the plaquette data, 
\beq
{\Delta \over T^4} = 6 N_t^4 \, {\partial \beta \over \partial {\rm log} a} 
\ \delta  P(\beta,T), \qquad \qquad \qquad
{p(T) \over T^4} - {p(T_0) \over T_0^4} = 6 N_t^4 \int_{\beta_0}^\beta d\beta
    \ \delta P(\beta,T)
\label{eq.eos}
\eeq
where $\delta P(\beta, T) = P(\beta, T) - P(\beta, T=0)$ 
is the difference in the plaquette observables between the finte 
temperature lattice and the corresponding zero temperature 
lattice, calculated at the coupling $\beta$.
$T_0$ is some reference temperature. 
$\partial \beta / \partial {\rm log} a$ is related to the beta function.
We use a nonperturbative estimate for the beta function, using the scaling
of $\tc$ from \cite{scaling}. Details of the beta function used by us can 
be found in \cite{thermo}. 

In the next section we discuss the latent heat of the deconfinement 
transition for the theory with N=3,4,6 number of colors. 
In Sec. \ref{sec.eos} we study the equation of state of
the SU(N) gluon plasma. A comparison of the energy density and pressure
for SU(3,4,6) gives us an idea of the size of the $1/{\rm N}^2$ corrections.
In the final section we discuss the physical implications of our
results.

SU(N) gauge theories have been investigated previously on the
lattice. In particular, the latent heat of the transition has been studied
in Ref. \cite{gavai,teper}, while the equation of state has been studied 
in Ref. \cite{barak,panero}. The focus of our work is in getting
results for the continuum limit. We use a nonperturbative beta
function, and multiple lattice spacings at
each temperature, for this purpose. We have also employed several
spatial volumes at each lattice spacing, to reach the thermodynamic
limit. Details of our work, including the cutoff and volume
dependence, can be obtained in Ref. \cite{thermo}. The results
reported here are our estimates of the infinite volume, continuum results.
An earlier version of our study, which used the two-loop beta
function, and had measurements at fewer temperatures, was 
presented in the previous year's conference \cite{lat09}. 

\section{Latent Heat}
\label{sec.lht}
SU(N) gauge theories have a first order deconfinement transition for N $\ge 
3$. A first order transition is characterized by a latent heat
associated with the transition. For the infinite volume system, the latent
heat can be defined as
\beq
   \frac{L_h}{T_c^4} = \displaystyle\lim_{\delta T\to0}
       \left(\frac{\epsilon(T_c+\delta T)}{T_c^4} 
            - \frac{\epsilon(T_c-\delta T)}{T_c^4}\right)
       = \displaystyle\lim_{\delta T\to0} \left(\frac{\Delta(T_c+\delta T)}
         {T_c^4} - \frac{\Delta(T_c-\delta T)}{T_c^4}\right)             
\label{eq.2state} \eeq
where $\epsilon(T_c \pm \delta T)$ are the energy densities of the
confined and deconfined phases, respectively, and the second equality 
follows from the fact that pressure is continuous across the
transition. 

\begin{figure}[htb]
\centerline{\includegraphics[width=.45\textwidth]
{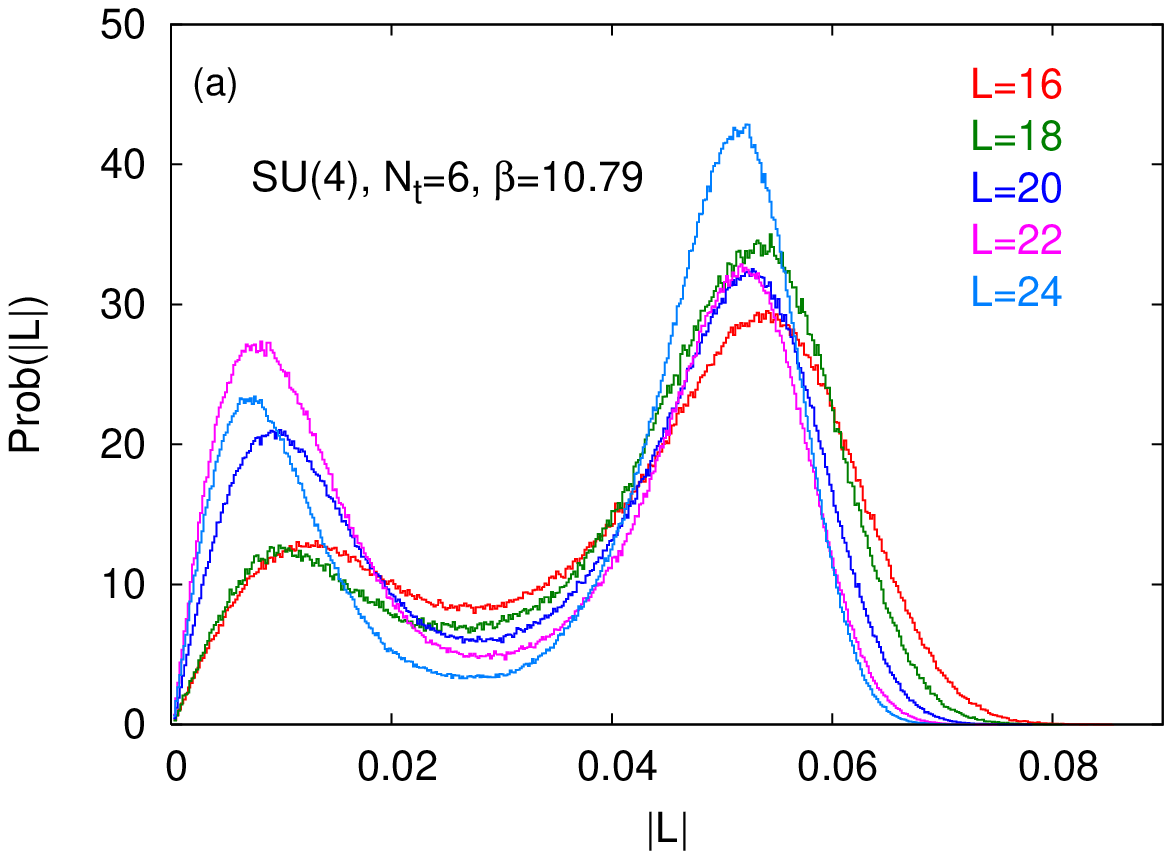}
\includegraphics[width=.45\textwidth]{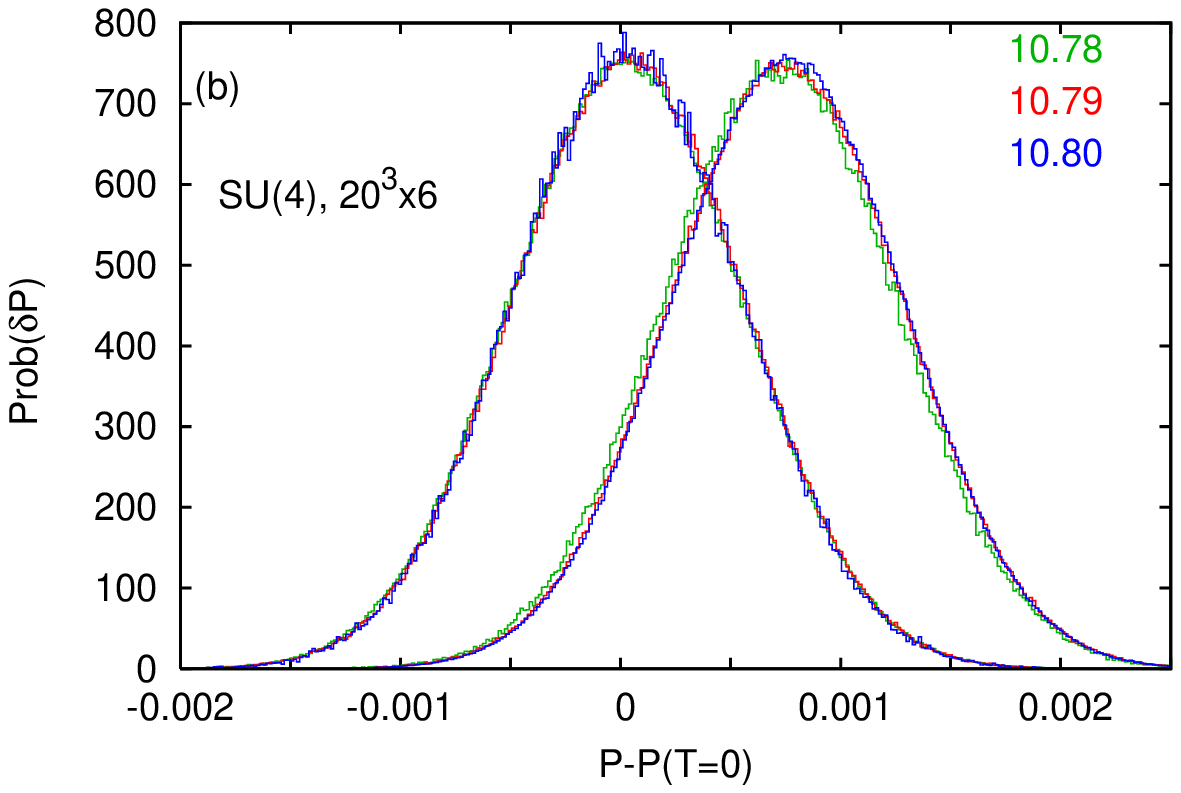}}
\caption{(a) Distribution of the Polyakov loop, |L|, in the transition
  region for SU(4) gauge theory. (b) Distribution of the plaquette, 
  $P(T)-P(T=0)$, for the cold and hot phases, identified by
  configurations with $|L| < L_c$ and $|L| > L_h$, respectively.}
\label{fig.lht}
\end{figure}

For a finite volume system, as is necessarily used in a numerical 
lattice computation, one cannot get a separation of the phases, and so,
a straightforward $\delta T \to 0$ limit will not work. 
We use the following method to extract the latent heat \cite{thermo}. 
Since the confined and the deconfined phases are resolved by the Polyakov 
loop $|L|$, we identify the configurations at the transition point
with $|L| < L_c$ as being in the 
confined phase and those with  $|L| > L_h$ as being in the deconfined phase,
where $L_c$ and $L_h$ are suitably chosen values. In order for the
procedure to be meaningful and not too sensitive on the choice of
$L_h$ and $L_c$, it is important that $|L|$ shows a two-peak
structure. Figure \ref{fig.lht} (a) shows the distribution of $|L|$ in
the transition regime of SU(4) gauge theory. $L_c$ and $L_h$ are chosen
at the valley (see Ref. \cite{thermo} for details). The distribution of
$\delta P(T)$ for the two phases defined by the $|L|$ cut is quite
stable, as shown in Fig. \ref{fig.lht} (b): for values of the coupling 
$\beta$ in the transition regime, the histograms for $\delta P(T)$ for the two
phases defined by the $|L|$ cut essentially coincide.

From the histograms of $\delta P(T)$ for the two phases at $\tc$, 
the latent heat can be calculated using Eq.(\ref{eq.2state}) and
Eq.(\ref{eq.eos}):
\beq
\frac{L_h}{\tc^4} \ = \ 6 N_t^4 \, \frac{\partial \beta}{\partial {\rm log} a}
\, \left( P(T_c^+) - P(T_c^-) \right)
\label{eq.lht} \eeq
where, as mentioned in the introduction, $\partial \beta / \partial 
{\rm log} a$ is evaluated from the nonperturbative scale-setting in 
\cite{scaling}. We find that the 
finite volume effects in the latent heat are rather large for SU(3), where 
one needs an aspect ratio $\zeta = LT > 6$ to reach the thermodynamic limit. 
On the other hand, for SU(4) and SU(6) gauge theories an aspect ratio 
$\zeta \ge 3$ suffices to reach the thermodynamic limit. This is probably 
related to the large correlation length of the Polyakov loop 
for SU(3) gauge theory near the 
transition. Our estimates for the thermodynamic limits of the latent heat 
for the theories with N = 3, 4 and 6 are shown in Table \ref{tbl.lht}.
To estimate the sensitivity of the results on the choices of $L_{c,h}$, we 
vary $L_h - L_c$ by $\pm 20 \%$; the corresponding change in the latent 
heat is shown as a systematic error in the table.

\begin{table}[htb]
\begin{center} 
\caption{Latent heat for SU(N) gauge theories with N=3, 4 and 6. The numbers
in brackets are the errors on the least significant digits; the first one is
statistical and the second, systematic.}
\label{tbl.lht}
\bigskip
\begin{tabular}{c|ccc}
\hline
N & 3 & 4 & 6 \\
$L_h / T_c^4$ & 1.67(4)(4) & 4.32(6)(6) & 11.93(34)(5) \\
\hline
\end{tabular} 
\end{center} \end{table}

Our results for the latent heat are in good agreement with 
Ref. \cite{teper}, within the larger
uncertainties of that study. The values obtained by Ref. \cite{gavai}
are somewhat higher. On going from N=3 to 4, the latent heat is seen to 
scale faster than the Casimir $d_A = {\rm N}^2-1$, the 
dimensionality of the adjoint
representation. On the other hand, the scaling 
between N=4 and 6 is consistent with a Casimir scaling. Taking all the three 
values in the Table, one can get a good fit to the relation 
\beq
\frac{L_h}{d_A T_c^4} \ = \ 0.388(3) - \frac{1.61(4)}{{\rm N}^2}
\label{eq.lhtnc}
\eeq
where the errors are statistical only.
Eq.(\ref{eq.lhtnc}) indicates a large correction to
the leading term for the theory with three colors. 

\section{Equation of State}
\label{sec.eos}

\begin{figure}[htb]
\centerline{\includegraphics[width=.6\textwidth]{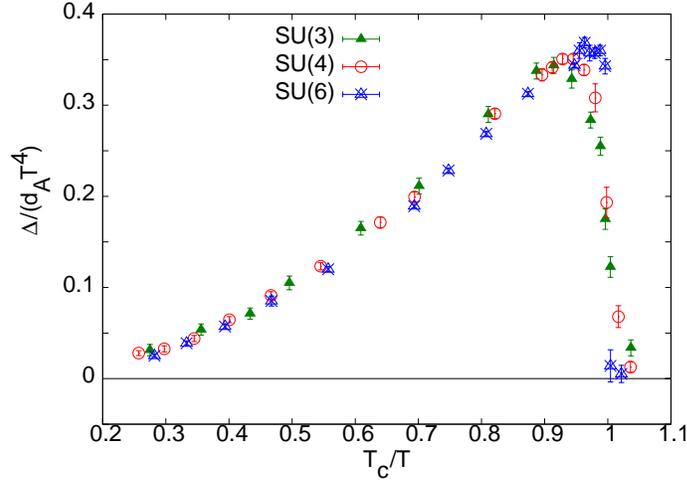}}
\caption{$\Delta/d_A T^4$ for SU(3-6) gauge theories, plotted
against $T_c/T$. For SU(3), we have used the plaquette data of Ref. 
\cite{boyd}.}
\label{fig.e3p}
\end{figure}

We first discuss $\Delta$, which is the trace of the energy-momentum
tensor and therefore is a measure of conformal symmetry
breaking. Fig. \ref{fig.e3p} shows $\Delta$ for SU(N) with
different number of colors, measured at different temperatures, and 
normalised by $d_A T^4$. 

As the figure shows, $\Delta$ scales nicely with $d_A$ except very
close to $\tc$. The size of the $1/{\rm N}^2$ correction is smaller than the 
statistical accuracy of our data, for $T > 1.25 \tc$. Closer to $\tc$
we see a deviation from scaling with $d_A$. The peak of $\Delta$ is
seen to become higher and move closer to $\tc$ with increasing N. 
At higher temperatures, $\Delta$ scales approximately like $T^2$ 
\cite{pisarski}, for the temperature range investigated by us \cite{thermo}.

The pressure was calculated using the so-called integral method
\cite{boyd}, Eq.(\ref{eq.eos}), which determines $p(T)$ in terms of
pressure at some reference temperature $p(T_0)$. We find that $p(T) \sim 0$
within our errors till temperatures $\sim 0.9 \tc$. This is probably
related to the fact that glueballs are much heavier than $\tc$, and
therefore are not excited substantially except very close to $\tc$.
We therefore evaluate $p/T^4$ by taking $\beta_0(T < 0.8\tc)$ as the
lower limit of the integral, with the knowledge that the omitted
additive constant is smaller than the statistical error. 

\begin{figure}[htb]
\centerline{\includegraphics[width=.5\textwidth]{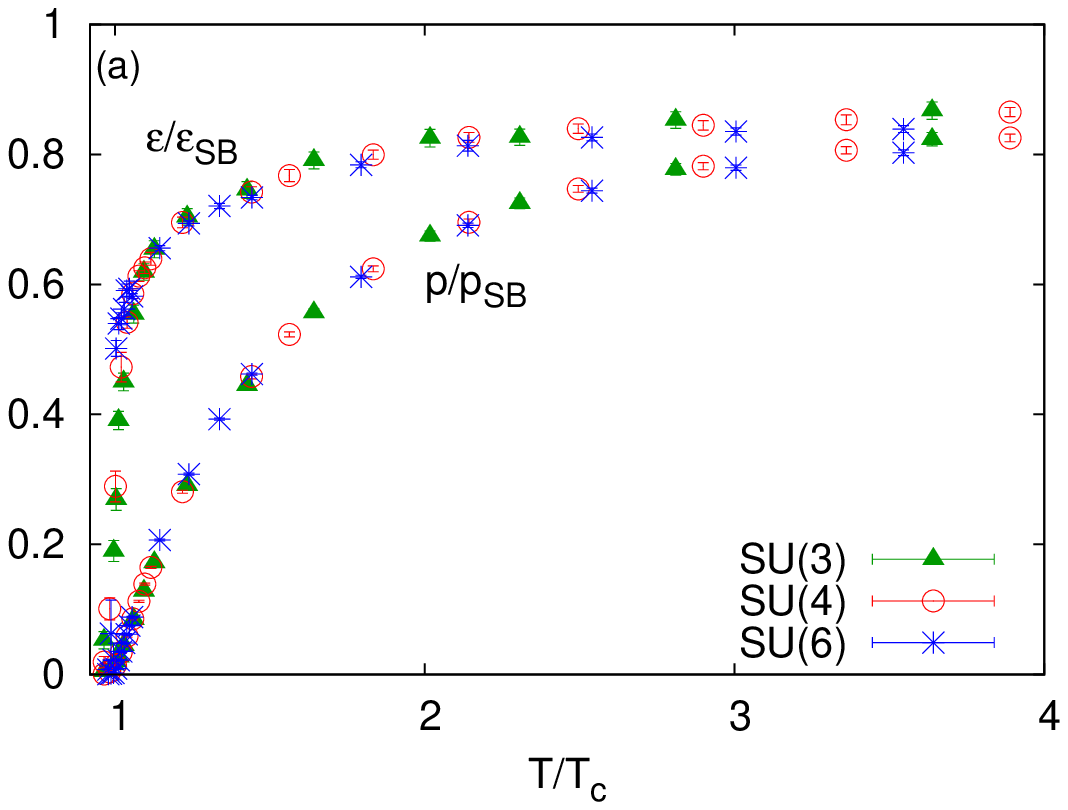}
\includegraphics[width=.5\textwidth]{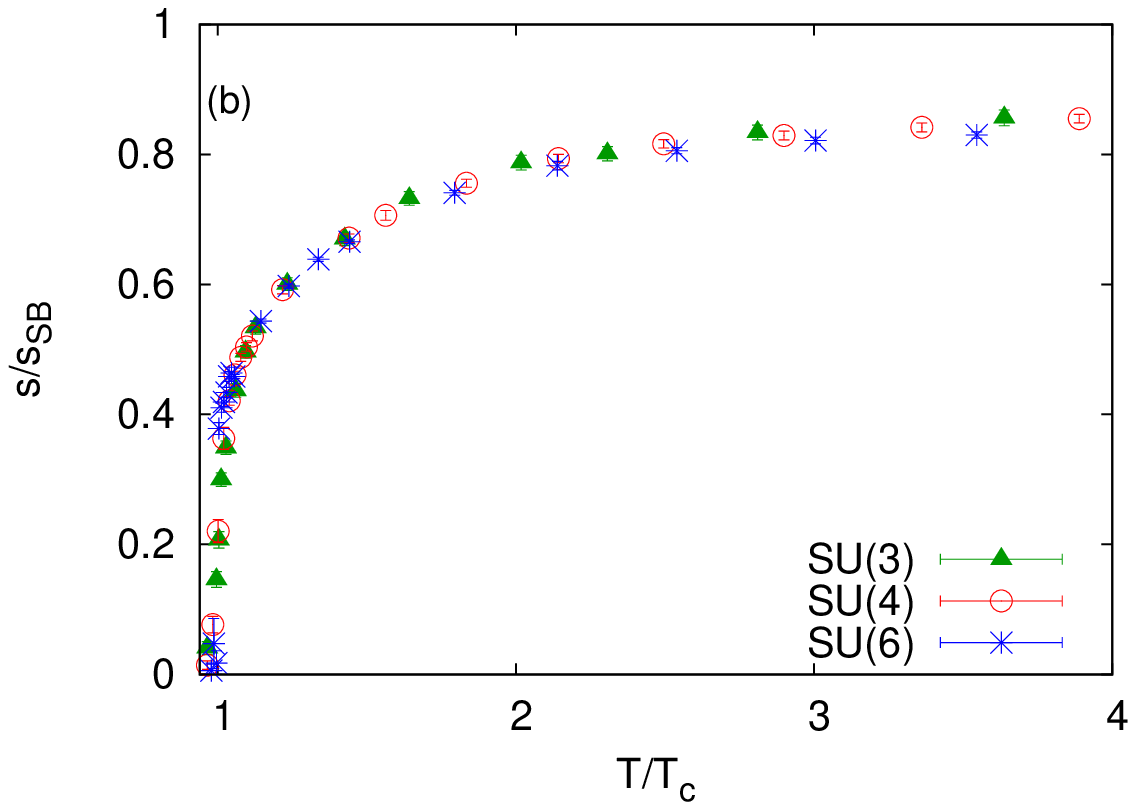}}
\caption{(a) The pressure and energy density for SU(N) gauge theories with 
N=3,4 and 6, normalized to their Stefan-Boltzman values. (b) The same for 
the entropy density.}
\label{fig.thermo}
\end{figure}

Other bulk thermodynamic quantities can be obtained from a knowledge of $p(T)$
and $\Delta(T)$. Figure \ref{fig.thermo} shows a summary of our
results for energy density, pressure and entropy density for SU(N) gauge 
theories, normalized to their Stefan-Boltzman
limits. The Stefan-Boltzman values used are the results calculated for
the free theory on the lattice, for the integral method \cite{engels}:
\beq
   \frac{\epsilon_{\stbo}}{T^4} = 3\,\frac{p_\stbo}{T^4}
        = \frac{\pi^2d_A}{15}G(N_t) \qquad{\rm where}\qquad
   G(N_t) = 1 + \frac{8\pi^2}{21} \;\frac1{N_t^2} + \cdots
\label{eq.sb}\eeq
As the figures show, even at temperatures close to 4
$\tc$, one reaches $\sim 85 \%$ or less of the Stefan-Boltzman value. 
The figures also 
show that, plotted as a function of temperature, the size of the $1/{\rm N}^2$ 
corrections is small in all the three observables, except for very close to 
$\tc$ where $1/{\rm N}^2$ corrections are visible in the energy density and 
entropy, as expected from the sizable $1/{\rm N}^2$ correction in the 
latent heat, Eq. (\ref{eq.lhtnc}). 

\section{Summary and Discussions:}
\label{sec.summary}
We have presented results for bulk thermodynamic quantities for SU(N)
gauge theories with N = 3, 4 and 6. Our results for latent heat of
the transition and its dependence on N are summarized in Table
\ref{tbl.lht} and Eq. (\ref{eq.lhtnc}). We find that the latent heat
for the theory with three colors is substantially less than that with 
larger number of colors: the size of the leading correction in
Eq. (\ref{eq.lhtnc}) is substantial for N=3.

\begin{figure}[htb]
\centerline{\includegraphics[width=.5\textwidth]{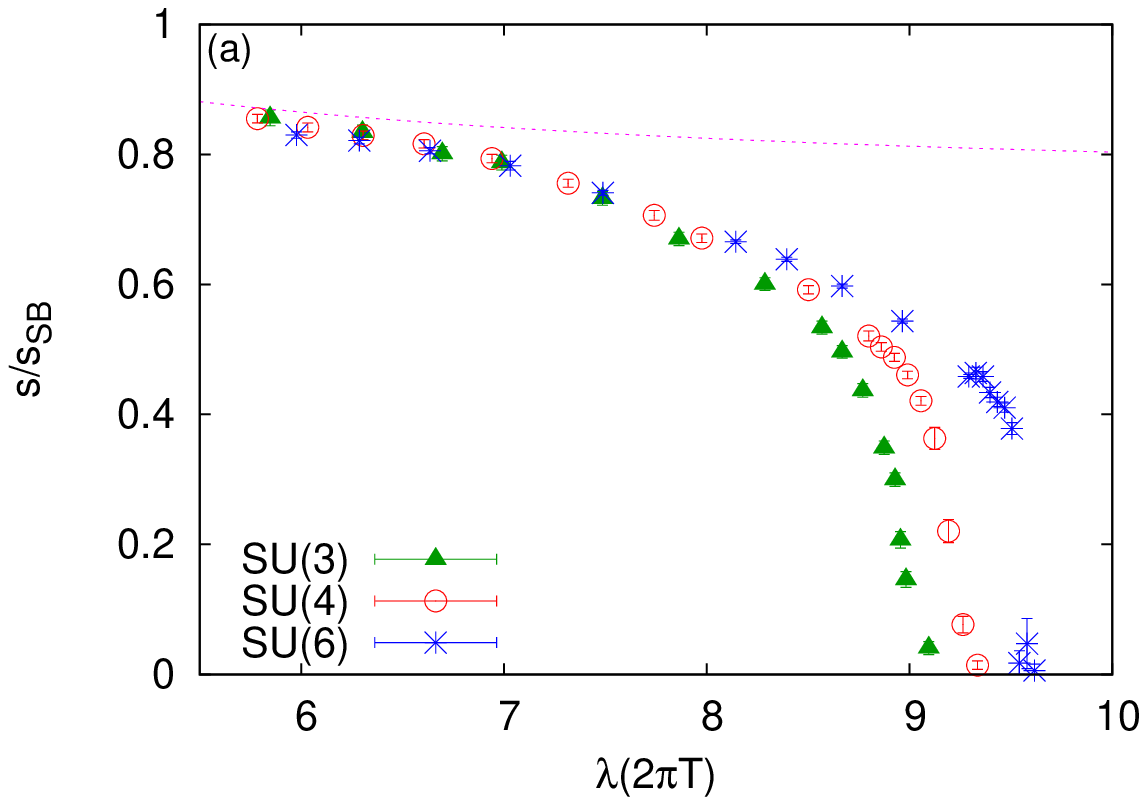}
\includegraphics[width=.5\textwidth]{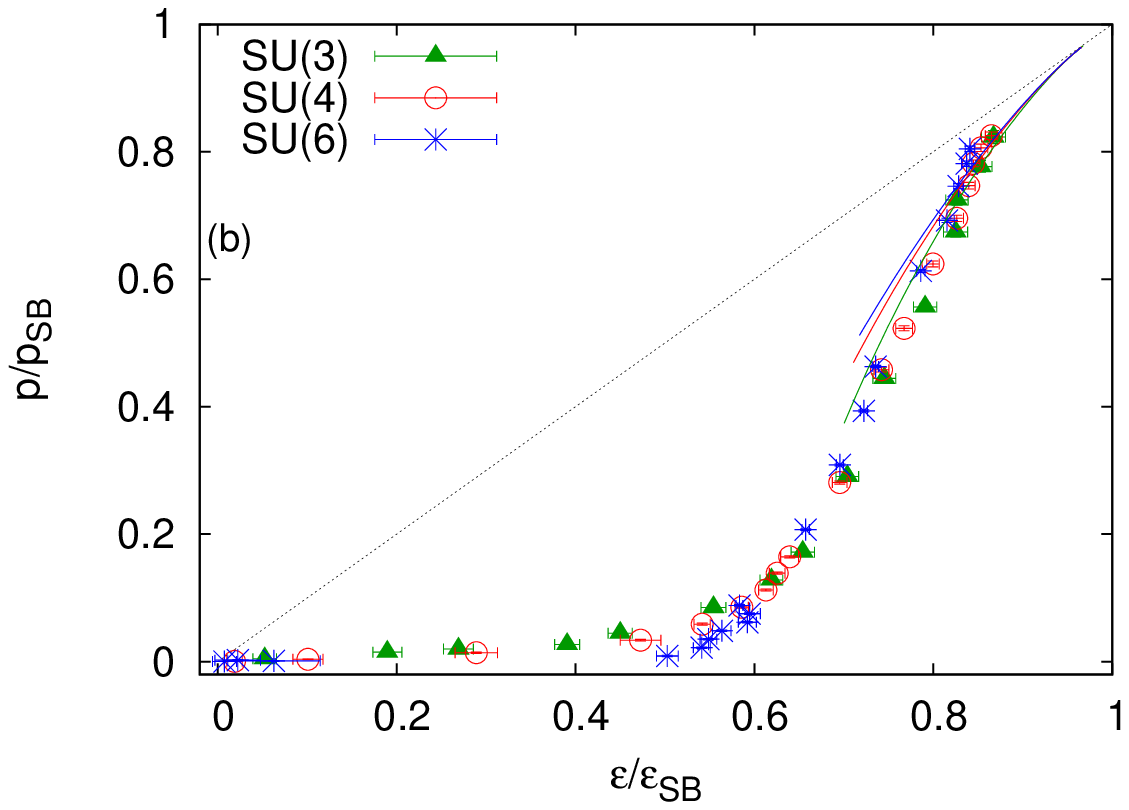}}
\caption{(a) Entropy density, in units of the Stefan-Boltzman value, 
  shown as a function of the 'tHooft coupling. The pink dotted line is the
  result for the $\mathcal{N} = 4$ supersymmetric SU(N) theory for
  large N \cite{gubser}. (b) $p/p_{\stbo}$ shown as a
  function of $\epsilon/\epsilon_{\stbo}$. The diagonal line is the line
  of conformal theories while, by construction, the Stefan-Boltzman
  limit is the point (1,1). The weak coupling results for the
  different theories are also shown.}
\label{fig.conformal}
\end{figure}

The bulk thermodynamic quantities are shown in Fig. \ref{fig.thermo}.
It is found that even at temperatures close to 4 $\tc$, one reaches 
only $\sim 85 \%$ of the Stefan-Boltzman value. We also see that 
except very close to $\tc$, the size of the leading $1/{\rm N}^2$
correction is small, when we compare the bulk thermodynamic
quantities measured at the same value of $T/T_c$. Interestingly,  
the scaling with the number of colors is better when looked at as
function of $T/T_c$ than when considered as function of the 't Hooft
coupling. Figure \ref{fig.conformal} (a) shows the entropy density
plotted against the 't Hooft coupling, defined through the V scheme
and evaluated at the scale $2 \pi T$, where the running is done
through our nonperturbative beta function. Considerable $1/{\rm N}^2$
correction is now seen, in particular for $T < 1.5 \tc$. Of course, 
the scaling with $\lambda$ will be different depending on the scheme
chosen to define the coupling. The figure also shows $s/s_{\stbo}$ for the
$\mathcal{N}$ = 4 supersymmetric theory \cite{gubser}. This theory is seen 
to have deviations from the Stefan-Boltzmann value of the same size as pure
SU(N) theory at the highest temperatures, but a very different dependence
on $\lambda$. 

The large deviation of the thermodynamic quantities from the 
Stefan-Boltzmann value even at 4 $\tc$, and the fact that
some strongly coupled conformal field theories show similar deviations
from Stefan-Boltzmann limit, have been used in the
literature to speculate about a strongly coupled, conformal regime in 
pure SU(N) gauge theories. The thermodynamics results can be used to
investigate the feasibility of such a phase. Following
Ref. \cite{swagato}, in Fig. \ref{fig.conformal} (b) we plot the energy
density vs. pressure, normalized by the corresponding Stefan-Boltzmann 
values. By construction the point at (1,1) is the Stefan-Boltzmann
limit, while the diagonal line denotes conformality. Also shown are
the weak coupling lines for the theories with the different number of
colors \cite{mikko}. We find that for the temperature regime of interest
to RHIC, $\sim 2 \tc$, the theory is far from conformal. At higher
temperatures, the theory approaches conformality, but it is closer
to the weak coupling line than the conformal line, indicating the 
absence of a strongly coupled, conformal phase in the SU(N) gluon plasma.

The computations were
carried out on the workstation farm of the department of theoretical
physics, TIFR. We thank Ajay Salve for technical support.

\end{document}